\documentclass[a4paper]{article}

\usepackage{INTERSPEECH2020}

\newcommand{\be}{\begin{equation}}
\newcommand{\ee}{\end{equation}}

\title{Improved Noisy Student Training for Automatic Speech Recognition}
\name{Daniel S. Park, Yu Zhang, Ye Jia, Wei Han, Chung-Cheng Chiu,
Bo Li, Yonghui Wu and Quoc V. Le}
\address{
  Google Inc., U.S.A.
 }
\email{\{danielspark, ngyuzh, jiaye, weihan, chungchengc, boboli, yonghui, qvl\}@google.com}

\begin{document}

\maketitle
\begin{abstract}
Recently, a semi-supervised learning method known as ``noisy student training" has been shown to improve image classification performance of deep networks significantly. Noisy student training is an iterative self-training method that leverages augmentation to improve network performance. In this work, we adapt and improve noisy student training for automatic speech recognition, employing (adaptive) SpecAugment as the augmentation method. We find effective methods to filter, balance and augment the data generated in between self-training iterations. By doing so, we are able to obtain word error rates (WERs) 4.2\%/8.6\% on the clean/noisy LibriSpeech test sets by only using the clean 100h subset of LibriSpeech as the supervised set and the rest (860h) as the unlabeled set. Furthermore, we are able to achieve WERs 1.7\%/3.4\% on the clean/noisy LibriSpeech test sets by using the unlab-60k subset of LibriLight as the unlabeled set for LibriSpeech 960h. We are thus able to improve upon the previous state-of-the-art clean/noisy test WERs achieved on LibriSpeech 100h (4.74\%/12.20\%) and LibriSpeech (1.9\%/4.1\%).
\end{abstract}

\section{Introduction}

We aim to improve semi-supervised learning in automatic speech recognition (ASR) by adapting a method proven successful for image classification that we refer to as ``noisy student training" (NST) \cite{noisystudent}. 
Noisy student training is an iterative self-training method that makes use of unlabeled data to improve accuracy. In noisy student training, a series of models are trained in succession, such that for each model, the preceding model in the series serves as a teacher model on the unlabeled portion of the dataset. The distinguishing feature of noisy student training is the exploitation of augmentation, where the teacher produces quality labels by reading in clean input, while the student is forced to reproduce those labels with heavily augmented input features. To ensure the integrity of the labels provided by the teacher, the data generated by the teacher is typically filtered by removing low confidence examples, and balanced by matching the labeled distribution.

We adapt noisy student training for the purpose of speech recognition by introducing the following measures:
\begin{enumerate}
\item We employ (adaptive) SpecAugment \cite{specaugment, largespecaugment}, an augmentation method for ASR that directly acts on the spectrogram of the input audio, for noisy student training.
\item We use shallow fusion with a language model (LM) \cite{shallowfusion} on the teacher network to generate better transcripts for the student network to train on.
\item We propose a \textbf{normalized filtering score} for transcripts generated by teacher networks given as a function of the fusion score and number of tokens.
\item We use a variant of sub-modular sampling \cite{shinohara2014} to weigh the utterance-transcript pairs generated by the teacher network to balance the token statistics of the dataset to be passed on to the student.
\end{enumerate}
When the supervised performance of the task is low, we find {gradational self-training} over NST generations effective:
\begin{enumerate}
\item We use \textbf{gradational filtering}, where the filtering criterion is sytematically relaxed to grow the semi-supervised dataset as the model performance improves.
\item We introduce \textbf{gradational augmentation}, where augmentation strength is increased with NST iterations.
\end{enumerate}

We are able to achieve state-of-the-art performance on two ASR tasks by using noisy student training. On the first task, LibriSpeech 100-860, the clean 100h subset of LibriSpeech \cite{librispeech} is used as the labeled set and the remaining 860h of LibriSpeech is used as the unlabeled set. We view this as a low-supervised-performance task, the baseline word error rates (WERs) being 5.5\%/16.9\% on the test-clean/test-other sets of LibriSpeech. By employing a noisy student training pipeline with gradational filtering and augmentation, we are able to achieve test-clean/test-other WERs 4.2\%/8.6\% on this task.

The second task is LibriSpeech-LibriLight, where the entirety of LibriSpeech is used as the labeled set, and the unlab-60k subset of LibriLight \cite{librilight}, a large untranscribed speech dataset based on audio books, is used as the unlabeled set. The baseline model \cite{han2020CNN} is able to achieve test-clean/test-other WERs 1.9\%/4.1\% showing high supervised performance. For this set, we obtain WERs 1.7\%/3.4\% on the test-clean/test-other sets by noisy student training without gradational filtering or augmentation, as we do not find gradational self-training effective.

\subsection{Related Work}

Self-training, where a teacher model is used to generate labels for the unlabeled set the student can train on, is one of the earliest methods in semi-supervised learning (e.g.,~\cite{scudder1965probability,yarowsky1995unsupervised,riloff2003learning}). Self-training has a long history of being applied to ASR through numerous studies, e.g., \cite{Zavaliagkos98utilizinguntranscribed,Novotney2009,Thomas2013,li2019,kahn2019selftraining,synnaeve2019endtoend,parthasarathi2019,hsu2020selfsupervised}. Noise has been studied and shown to have positive effects in self-training for machine translation and image classification recently~\cite{noisystudent,he2020revisiting}, directly motivating our work.
Gradational growth of the labelled dataset in self-training has been used in \cite{Lamel00lightlysupervised}.

Another direction of semi-supervised learning research is to use consistency-based methods (e.g., \cite{xie2019unsupervised,zhai2019s4l,sohn2020fixmatch}). In this approach, a separate task/loss to (pre-)train the acoustic model (or a part thereof) is introduced to help the model learn a good representation of the input, e.g., \cite{hsu2018extracting, chung2018speech2vec, chung2019autoregressive, chrowski2019unsupervised, schneider2019wav2vec, baevski2019vqwav2vec, ling2019deep}.

Our work builds upon self-training and has three main contributions. First, we make use of new augmentation methods in speech to improve the task performance. Second, we introduce a normalized filtering score for shallow-fused models for filtering generated transcripts. Third, we apply gradational filtering where the filtering criterion for unlabeled data used for training is systematically relaxed with the self-training iterations.

\section{Noisy Student Training for ASR}
\label{s:nst}

The NST algorithm we propose assumes a labeled set $S$, an unlabeled set $U$ and a fixed LM trained on a separate text corpus. Then, NST generates a series of ASR models as follows:
\begin{enumerate}
\item Train $M_0$ on $S$ with SpecAugment. Set $M = M_0$.
\item Fuse $M$ with LM and measure performance.
\item Generate labeled dataset $M(U)$ with fused model.
\item Filter generated data $M(U)$ to obtain $f(M(U))$.
\item Balance filtered data $f(M(U))$ to obtain $b \cdot f(M(U))$.
\item Mix dataset $b \cdot f (M(U))$ and $S$. Use mixed dataset to train new model $M'$ with SpecAugment.
\item Set $M = M'$ and go to 2.
\end{enumerate}
We denote each cycle of the noisy student training a generation. We denote the filtered and balanced dataset $b \cdot f (M(U))$ the ``semi-supervised portion" of the training set. We now elaborate on the details of the presented components.
\smallskip

\noindent\textbf{SpecAugment:}
We use SpecAugment \cite{specaugment} to augment the input data at each step of NST. We employ adaptive time masking \cite{largespecaugment}, where the size of the time masks scales linearly with the length of the input utterance, depending on the task.
\smallskip

\noindent\textbf{Language Model Fusion: }
To produce better transcripts for the student, we shallow-fuse \cite{shallowfusion} the teacher networks with an LM trained on a fixed text corpus. We introduce a coverage penalty term \cite{coveragepenalty} in the fusion score with parameter $c$ for LAS models \cite{LAS}, while we use a constant non-blank reward $\rho$ to shift the logits produced by the LM for RNN-T \cite{RNNT} networks \cite{sak2015fast,zhang2020transformer}. These parameters are tuned at each generation via grid search to minimize the dev-set WERs. The transcripts of the unlabeled set are generated from the fused model using beam search.
\smallskip

\noindent\textbf{Filtering: }
Given that the generated transcripts were found based on a fusion score that does not have an interpretation as a log-probability, it is unclear what to use to judge the quality of the transcripts generated by a teacher network.

We introduce a filtering score given as a function of the shallow fusion score $\mathcal{S}$ and the token length $\ell$ of a transcript generated by the fused teacher model. The normalized filtering score is defined to be
\be\label{score}
s(\mathcal{S},\ell) = (\mathcal{S} - \mu \ell  - \beta) / (\sigma \sqrt{\ell}) \,,
\ee
with parameters $\mu$, $\beta$ and $\sigma$ fit on the dev-set. $\mu$, $\beta$ are fit via linear regression on the value pairs $(\ell_i, \mathcal{S}_i)$ of the transcripts generated from the dev-set utterances. $\sigma$ is obtained by computing the standard deviation of $(\mathcal{S}_i - \mu \ell_i -\beta) / \sqrt{\ell_i}$. These parameters are fit for each new generation of teacher models.

Using this score, we filter the transcript-utterance pairs generated by the trained models in a gradational manner, i.e., we lower the filtering cutoff $s_\text{cutoff}$ for which only the utterances with scores $s(\mathcal{S}, \ell) > s_\text{cutoff}$ are kept, as the self-training cycle is iterated. This strategy is expected to work best when the teacher model becomes markedly better with NST iterations.
\smallskip

\noindent\textbf{Balancing: }
The distribution of tokens of the transcripts of the generated set $f(M(U))$ can differ significantly from that of the supervised training set distribution. We may choose to weigh the samples in the newly generated dataset to bridge this gap.

We use sub-modular sampling \cite{shinohara2014} with the ``cost-benefit score" to balance the dataset. Sub-modular sampling is a method that samples (with replacement) a set of sentences from a sentence pool so that the token distribution of the sampled set is close to a target distribution. This is done by optimizing the KL divergence between the token distributions of the sampled set and the target distribution in a greedy way, where sentences are collected by batches of size $B$. Each batch is chosen from the pool of sentences by selecting the top-$B$ sentences in terms of cost-benefit, i.e., the decrease in KL divergence by adding that sentence to the current set of sampled transcripts, divided by its number of tokens. After a batch of samples have been collected and added to the sampled set, the KL divergence between the token distribution of assembled sentences and the target distribution is computed, and the process is iterated.

We impose additional constraints on this algorithm by capping the sampling multiplicity of a sentence by $2$  and requiring the total number of tokens in the sampled set to be bounded below by the total number of tokens in the supervised set. Furthermore, we take the batch size to be a tenth of the filtered dataset. We optimize the KL divergence with these restrictions to obtain the balanced set $b \cdot f (M(U))$ from $f(M(U))$.
\smallskip

\noindent\textbf{Mixing: }
We combine  $b \cdot f (M(U))$ with the supervised dataset $S$ for training using two different methods of mixing. In batch-wise mixing, we fix the ratio between supervised and semi-supervised samples in each training batch. In non-batch-wise mixing, data is uniformly sampled from both datasets.

\section{Experiments}

\subsection{LibriSpeech 100-860}
\label{s:100-860}
LibriSpeech 100-860 is a semi-supervised task where the clean 100h subset of LibriSpeech \cite{librispeech} is taken to be the supervised set, while the remaining 860h of audio is taken to be the unlabeled set. The unlabeled audio consists of 360h of clean data and 500h of noisy data. We tokenize the transcripts using a WPM model \cite{wpm} with vocabulary size 16k constructed from the clean 100h subset transcripts. We use 80-dimensional filter bank coefficients with delta and delta-delta acceleration as the input features. We use the model denoted LAS-6-1280 in \cite{specaugment} (see also \cite{lasbaseline}) as our acoustic model, which is a Listen, Attend and Spell network \cite{LAS} with a bi-directional LSTM encoder.

We train 6 generations of models numbered 0 to 5, where we count the baseline model trained with the supervised set as the zeroth generation. Each generation is trained with peak Adam learning rate of $0.001$ and batch size of 512 on 32 Google Cloud TPU chips for 10 days. The checkpoint to be used for LM fusion is chosen based on its dev-set WER. The choices made for the components of NST are as follows:

\noindent\textbf{SpecAugment:} For generation 0, we use two frequency masks with mask parameter ($F$) 27, two time masks with mask parameter ($T$) 40, and time warping with warp parameter ($W$) 40 \cite{specaugment}. At generations 2 and 4, we re-tuned the time mask parameter $T$ on a proxy task (with model LAS-4-1024 and schedule LB from \cite{specaugment}). The time masking parameter used for generations 0 and 1, 2 and 3, 4 and 5 were set to 40, 80 and 100, respectively.

\noindent\textbf{LM:} We use a 3-layer LSTM language model with width 4096 trained on the LibriSpeech LM corpus tokenized with the 16k WPM. The LM has word-level perplexity 63.9 on the dev-set.

\noindent\textbf{Filtering:} We apply gradational filtering from generation 1 to 5 with cutoffs, 1, 0.5, 0, $-1$ and $-\infty$. We apply filtering separately to the clean-360h and other-500h audio, i.e., the scoring coefficients are fit separately on dev-clean and dev-other, and applied to clean-360h and other-500h respectively.

\noindent\textbf{Other:} Neither balancing or batch-wise mixing is used.

The performance of the fused models through the generations are plotted in figure \ref{f:librigens}. Our best trained model is the generation-4 model. We compare the performance of our baseline model, as well as the generation-4 model before and after LM fusion, against other results in the literature in table \ref{t:100sota}.

\begin{table}[h!]
  \vskip -0.05in
  \caption{LibriSpeech 100h WERs (\%).}
  \label{t:100sota}
  \vskip -0.05in
  \centering
  \small
  \resizebox{\columnwidth}{!}{%
  \begin{tabular}{lcccc}
    \toprule
    \bfseries Method & \multicolumn{2}{c}{\bfseries Dev} & \multicolumn{2}{c}{\bfseries Test} \\
    \cmidrule(r){2-3} \cmidrule(r){4-5}
     & \bfseries clean & \bfseries other & \bfseries clean & \bfseries other \\
    \midrule
    \bfseries Supervised \\
    \quad  L\"{u}scher et al., (2019) \cite{luscher2019interspeech}
    & 5.0 & 19.5 & 5.8 & 18.6 \\
    \quad Kahn et al., (2019) \cite{kahn2019selftraining}
    & 7.78 & 28.15 & 8.06 & 30.44 \\
    \quad Hsu et al., (2019) \cite{hsu2020selfsupervised}
    & 14.00 & 37.02 & 14.85 & 39.95 \\
    \quad Ling et al., (2019) \cite{ling2019deep}
    & & & 6.10 & 17.43 \\
    \midrule
    \bfseries Semi-supervised (w/ LibriSpeech 860h)\\
    \quad Kahn et al., (2019) \cite{kahn2019selftraining}
    & 5.41 & 18.95 & 5.79 & 20.11 \\
    \quad Hsu et al., (2019) \cite{hsu2020selfsupervised}
    & 5.39 & 14.89 & 5.78 & 16.27 \\
    \quad Ling et al., (2019) \cite{ling2019deep}
    & & & 4.74 & 12.20 \\
    \midrule
    \bfseries This Work \\
    \quad Baseline (LAS + SpecAugment) & 5.3 & 16.5 & 5.5 & 16.9 \\
    \qquad + NST before LM Fusion & 4.3 & 9.7 & 4.5 & 9.5 \\
    \qquad + NST with LM Fusion & \bfseries 3.9 & \bfseries 8.8 & \bfseries 4.2 & \bfseries 8.6 \\
    \bottomrule
  \end{tabular}
  }
  \vskip -0.1in
\end{table}

\subsection{LibriSpeech-LibriLight}
\label{s:ls-ll}
The entire LibriSpeech \cite{librispeech} training set is used as the supervised training set of this task, while the ``unlab-60k" subset of LibriLight \cite{librilight}, an unlabeled audio dataset derived from audio books, is used as the unlabeled set. We do not use the ``duplicate" set, nor any supervised subset of LibriLight. We segment the audio using a maximum expected segmentation length of 36 seconds, but filter further down to choose segments that have less than 22 seconds of audio. This results in producing  approximately a million utterances. The transcripts of the supervised set are tokenized using a WPM model with vocabulary size 1k constructed from the LibriSpeech training set transcripts.

\begin{figure}[h]
  \vskip -0.05in
  \centering
  \begin{tabular}{c}
  \includegraphics[width=0.95\linewidth]{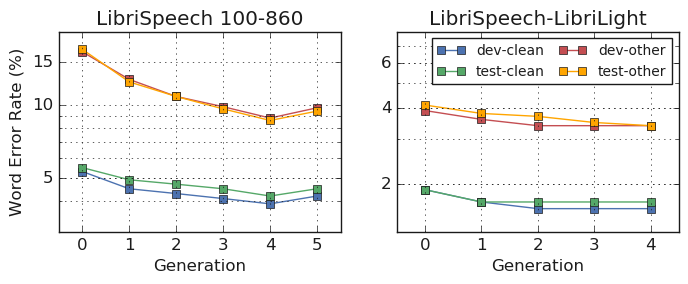}
  \end{tabular}
  \vskip - 0.05in
  \caption{WERs (\%) plotted against training generation.}
  \label{f:librigens}
  \vskip -0.1in
\end{figure}

We use 80-dimensional filter bank coefficients of the utterances as single-channel input features. We use a family of ContextNets \cite{han2020CNN}, which are RNN-T models \cite{RNNT} with deep CNN encoders, for this task. We use CN-$w$ to denote a model with a 23-convolutional-block encoder and 2-layer decoder with cell size 1280 and width scaling factor $w$ studied in \cite{han2020CNN}.

Five generations of models numbered 0 to 4, are trained, where the baseline model is taken to be the generation-zero model. The baseline ContextNet model has the same encoder as CN-2, but has a one-layer RNN decoder with dimension 640. Meanwhile, CN-$w$ with $w$=$1.25$, $1.75$, $2.25$ and $2.5$ have been set to be the ASR model from generation 1 through 4. Each generation is trained on 128 Google Cloud TPU chips for 1-3 days. The checkpoint of the model to be used for LM fusion is chosen based on the dev-set WER. The choices for the NST pipeline have been set as follows:

\noindent\textbf{SpecAugment:} We use SpecAugment with adaptive time mask size \cite{specaugment, largespecaugment}. We use two frequency masks with mask parameter ($F$) 27, and ten time masks with maximum time-mask ratio ($p_S$) 0.05, where the maximum-size of the time mask is set to $p_S$ times the length of the utterance. Time warping is not used.

\noindent\textbf{LM:} We use a 3-layer LSTM LM with width 4096 trained on the LibriSpeech LM corpus with the LibriSpeech 960h transcripts added, tokenized with the 1k WPM. The LM has word-level perplexity 68.3 on the dev-set transcripts.

\noindent\textbf{Other:} We do not apply filtering, while balancing is applied with the parameters detailed in section \ref{s:nst}. We use batch-wise mixing, and gradually increase the portion of the semi-supervised utterances---the ratio between the supervised and semi-supervised utterances are raised from 4:6 at generations 1, 2 to 3:7 at generation 3 to 2:8 at generation 4.

We have plotted the generational performance of the fused models in figure \ref{f:librigens}. Our best trained model is the generation-4 model whose performance is presented in table \ref{t:sota}.

\begin{table}[h!]
  \vskip -0.05in
  \caption{LibriSpeech 960h WERs (\%).}
  \label{t:sota}
  \vskip -0.05in
  \centering
  \small
  \resizebox{\columnwidth}{!}{%
  \begin{tabular}{lcccc}
    \toprule
    \bfseries Method & \multicolumn{2}{c}{\bfseries Dev} & \multicolumn{2}{c}{\bfseries Test} \\
    \cmidrule(r){2-3} \cmidrule(r){4-5}
     & \bfseries clean & \bfseries other & \bfseries clean & \bfseries other \\
    \midrule
    \bfseries Supervised \\
    \quad  Synnaeve et al., (2019) \cite{synnaeve2019endtoend}
    & 2.10 & 4.79 & 2.33 & 5.17 \\
    \quad Zhang et al., (2020) \cite{zhang2020transformer}
    & & & 2.0 & 4.6 \\
    \quad Han et al., (2020) \cite{han2020CNN} $^\dagger$
    & 1.9 & 3.9 & 1.9 & 4.1 \\
    \midrule
    \bfseries Semi-supervised \\
    \quad  Synnaeve et al., (2019) \cite{synnaeve2019endtoend}
    & 2.00 & 3.65 & 2.09 & 4.11 \\
    \midrule
    \bfseries This Work (with baseline $\dagger$)\\
    \quad ContextNet + NST before LM Fusion
    & \bfseries 1.6 & 3.7 & \bfseries 1.7 & 3.7 \\
    \quad ContextNet + NST after LM Fusion
    & \bfseries 1.6 & \bfseries 3.4 & \bfseries 1.7 & \bfseries 3.4 \\
    \bottomrule
  \end{tabular}
  }
  \vskip -0.1in
\end{table}

\section{Discussion}
\label{s:discussion}

\subsection{LibriSpeech 100-860}

\begin{figure}[h]
  \vskip -0.05in
  \centering
  \begin{tabular}{c}
  \includegraphics[width=0.95\linewidth]{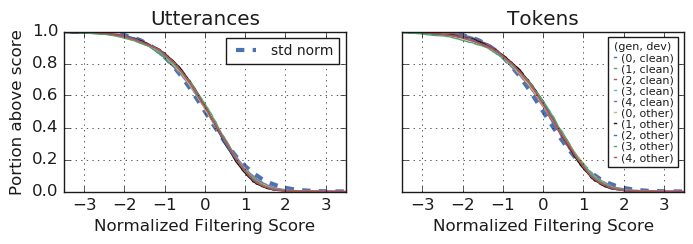}
  \end{tabular}
  \vskip -0.05in
  \caption{Portion of utterances/tokens above filtering score. Solid curves are indexed by generation and filtered dev-set.}
  \label{f:scoredist}
  \vskip -0.1in
\end{figure}

\noindent\textbf{Gradational Filtration:} As noted in section \ref{s:100-860}, for LibriSpeech 100-860, we gradually relax the filtering criterion for the semi-supervised dataset based on their normalized filtering score. Recall that we generate transcripts for the dev-clean and dev-other sets using the generation 0 to 4 LibriSpeech 100-860 models to fit the filtering score parameters. We plot the portion of utterances and their tokens in these datasets that are above a given filtering score in figure \ref{f:scoredist}. The thick dotted line is the cumulative density of the standard normal distribution we have plotted as a reference, while the ten solid curves come from transcripts generated at each generation from the two dev-sets. The distribution of utterances are remarkably stable throughout generations, the solid curves virtually lying on top of each other.

\begin{figure}[h]
  \vskip -0.05in
  \centering
  \begin{tabular}{c}
  \includegraphics[width=0.95\linewidth]{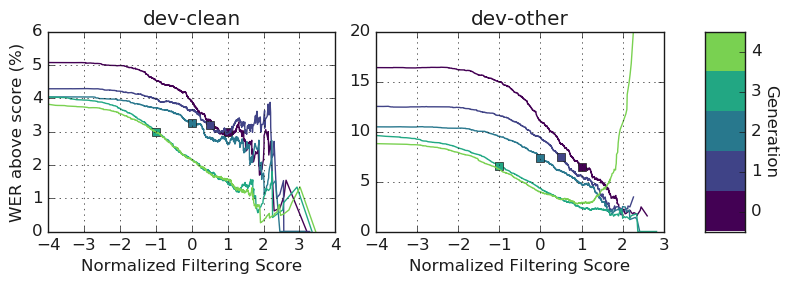}
  \end{tabular}
  \vskip -0.05in
  \caption{WER above filtering score for LibriSpeech 100-860 for dev-clean and dev-other with generation 0-4 models.}
  \label{f:WERgens}
  \vskip -0.1in
\end{figure}

We have plotted the WER of utterances above a given filtering score in figure \ref{f:WERgens} for each generation. The score reasonably ranks the quality of the generated dev-set transcripts, except at generation 4, where the model generates a blank transcript to a long utterance, which makes the cumulative WER become large at high scores. We have marked the score used for filtering at each generation. The filtered dev-set WERs are comparable throughout the generations, while the size of the filtered dataset grows. This is the intended effect of gradational filtration. 

\begin{figure}[h]
  \centering
  \vskip -0.05in
  \begin{tabular}{c}
  \includegraphics[width=0.95\linewidth]{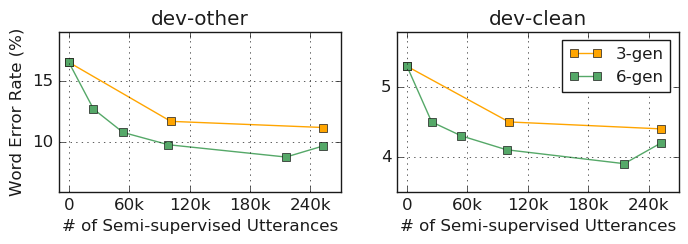}
  \end{tabular}
  \vskip -0.05in
  \caption{WER (\%) vs semi-supervised dataset size for models in 6 and 3 generation noisy student training pipelines.}
  \label{f:evolutionspeed}
  \vskip -0.1in
\end{figure}

We find that slow evolution improves performance in the range we have explored. We have run an experiment where we evolve the model ``faster," by training two generations in addition to the baseline, using filtering cutoffs 0 and $-\infty$. We plot the evolution of dev-set WERs against the semi-supervised dataset size in figure \ref{f:evolutionspeed}. Models in the 6-gen pipeline outperforms those in the 3-gen pipeline at comparable dataset sizes.

\noindent\textbf{Augmentation:} The increase of time mask size used for training at generations 2 and 4 is based on proxy task evaluation. In the proxy task, LAS-4-1024 models \cite{specaugment,lasbaseline}, which have about 1/3 of the parameters of LAS-6-1280, are trained for 100k steps with the combined training set at generations 2 and 4 with varying time mask size. The time mask size has been chosen based on the final WERs tabulated in table \ref{t:augtune}, where we prioritized performance on the clean set. The performances are measured without LM fusion. We find that a stronger augmentation is favored, as the dataset grows with gradational filtering.

\begin{table}[h!]
  \vskip -0.05in
  \caption{Proxy task WER (\%) for LibriSpeech 100-860.}
  \label{t:augtune}
  \vskip -0.05in
  \centering
  \small
  \resizebox{0.9\columnwidth}{!}{%
  \begin{tabular}{ccrrrrr}
    \toprule
    \multicolumn{2}{c}{Time Mask Size ($T$)} & 40 & 60 & 80 & 100 & 120 \\
    \midrule
    Gen 2 & dev-clean & 5.7 & 5.6 & 5.5 & 9.2 & -  \\
    & dev-other & 15.0 & 14.5 & 14.7 & 17.8 & -  \\
    \midrule
    Gen 4 & dev-clean & - & 4.8 & 5.0 & 4.7 & 4.7 \\
    & dev-other & - & 11.7 & 11.7 & 11.4 & 11.7 \\
    \bottomrule
  \end{tabular}
  }
  \vskip -0.05in
\end{table}

\begin{figure}[h]
  \centering
  \begin{tabular}{c}
  \includegraphics[width=0.95\linewidth]{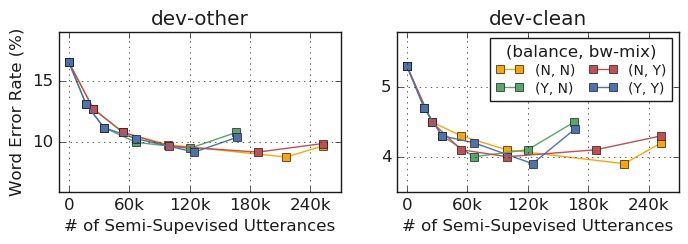}
  \end{tabular}
  \vskip -0.05in
  \caption{WERs plotted against the size of the semi-supervised dataset for combinations of balancing and batch-wise mixing.}
  \label{f:bm}
  \vskip -0.15in
\end{figure}

\noindent\textbf{Mixing and Balancing:}  We conduct control experiments with batch-wise mixing and balancing for LibriSpeech 100-860. We have experimented with four combinations of balancing and 1:1 batch-wise mixing for the LibriSpeech 100-860 task, the results of which we present in figure \ref{f:bm}. Balancing cuts down the semi-supervised dataset size at each generation to about a half, diminishing the performance of the trained network.

\subsection{LibriSpeech-LibriLight}

We present control experiments for the LibriSpeech-LibriLight task. A slightly inferior baseline with model CN-1.25 with dev-clean/other WERs 1.8\%/4.2\% and test-clean/other WERs 1.9\%/4.4\% has been used for these experiments.
\smallskip

\noindent\textbf{Balancing and Filtering:} Combinations of filtering with score 0 (which filters $\sim$ 50\% of the utterances) and balancing have been tested on the first generation of NST. The pre-fusion WERs are presented in table \ref{t:llcontrol}. In contrast with LibriSpeech 100-860, the reduction in the semi-supervised set size by balancing does not seem to outweigh the benefits of making the token distribution closer to that of the supervised set, due to the size of the LibriLight dataset. We have chosen to use balancing, but not to filter in our pipeline based on these experiments.

\begin{table}[h!]
  \vskip -0.05in
  \caption{WERs (\%) for balancing and filtering configurations.}
  \label{t:llcontrol}
  \vskip -0.05in
  \centering
  \small
  \resizebox{0.8\columnwidth}{!}{%
  \begin{tabular}{ccccc}
    \toprule
    (Bal, Filt) & (N, N) & (N, Y) & (Y, N) & (Y, Y) \\
    \midrule
    dev-clean & 1.8 & 1.8 & 1.7 & 1.8 \\
    dev-other & 4.5 & 4.4 & 4.3 & 4.3 \\
    \bottomrule
  \end{tabular}
  }
  \vskip -0.05in
\end{table}

\noindent\textbf{Gradational Self-Training:} We find that gradational filtration and augmentation are not as effective for this task as it was for LibriSpeech 100-860. We compare the performance of a network trained with gradational filtration over 5 generations with filtering scores $1$, $0.5$, $0$, $-1$ and $-\infty$ and one trained on the entire LibriLight dataset without filtering. We do not see the benefit of gradational filtration, as both models, after LM fusion have dev-clean/dev-other WERs 1.7\%/3.7\%, respectively. An explanation of this result is that the baseline model for these tasks is already extremely good, and the quality of filtered transcripts is not drastically better than the unfiltered transcripts. For example, the WER for the transcripts with score $> 1$ on the dev-set is 1.5\%, against 3.0\% of the unfiltered transcripts at generation 0. This should be compared to the $\sim$ 10\% difference in WER between the transcripts with score $> 1$ against the unfiltered transcripts generated from the dev-other utterances at generation 0 for LibriSpeech 100-860 (gen-0 curve in figure \ref{f:WERgens}).

\section{Conclusion}

We have adapted and improved noisy student training for ASR. We have employed SpecAugment, language model fusion and sub-modular sampling into the noisy student training pipeline to adapt it for speech recognition. We have also introduced a normalized filtering score which we use for gradational self-training. The gradational methods prove to be beneficial for the low-supervised-performance task LibriSpeech 100-860, while it does not show much effect for the high-supervised-performance task LibriSpeech-LibriLight. The various components introduced in this work were mixed and matched to improve the state-of-the-art performance of both tasks studied.
\smallskip

\noindent\textbf{Acknowledgements:} We thank William Chan, Zhehuai Chen, Mike Chrzanowski, Ekin Dogus Cubuk, Arun Narayanan, Sankaran Panchapagesan, Bhuvana Ramabhadran, June Shangguan and Barret Zoph for helpful discussions.

\bibliographystyle{IEEEtran}

\bibliography{mybib}

\end{document}